\documentclass[aps,prd,twocolumn,groupedaddress,nofootinbib,preprintnumbers,superscriptaddress]{revtex4-1}  
\usepackage{graphicx,mathtools,tabu,array}
\usepackage{epstopdf}
\usepackage{amsmath}
\usepackage{amsfonts}
\usepackage[colorlinks=true,citecolor=red,linkcolor=blue,breaklinks=true]{hyperref}
\usepackage{accents}
\usepackage[figure, table]{hypcap}
\usepackage{amssymb}
\usepackage{appendix}
\usepackage{comment}
\usepackage{bbold}
\usepackage{xcolor}
\usepackage{slashed}
\usepackage{subfigure}
\usepackage{setspace}
\usepackage{footnote}
\usepackage{multirow}
\usepackage{braket}
\usepackage[normalem]{ulem}
\usepackage[utf8]{inputenc}
\usepackage{mathrsfs}
\usepackage{longtable}
\usepackage{enumitem}
\usepackage{orcidlink}

\newcommand{\PreserveBackslash}[1]{\let\temp=\\#1\let\\=\temp}
\newcolumntype{C}[1]{>{\PreserveBackslash\centering}p{#1}}
\newcolumntype{R}[1]{>{\PreserveBackslash\raggedleft}p{#1}}
\newcolumntype{L}[1]{>{\PreserveBackslash\raggedright}p{#1}}

\LTcapwidth=\textwidth

\usepackage{url}
\usepackage{wrapfig}
\usepackage{braket}

 \def\be   {\begin{equation}}  
 \def\ee   {\end{equation}}
 \def\ba   {\begin{array}}     
  \def\ea   {\end{array}}
 \def\bea  {\begin{eqnarray}}  
  \def\eea  {\end{eqnarray}}
 \def\bean {\begin{eqnarray*}}  
 \def\eean {\end{eqnarray*}}

  \def\be {\beta}

\def\to {\rightarrow}

\newcommand{\cnub}{C$\nu$B}
\newcommand{\K}{\,\mathrm{K}}

\newcommand{\ms}{\,M_\odot}

\newcommand{\zmax}{{z_\mathrm{max}}}

\newcommand{\COM}[1]{\textcolor{black}{#1}}

\newcommand{\eV}{{\rm\ eV}}

\newcommand{\MeV}{{\rm\ MeV}}

\begin{document}

\title{Neutrino secret self-interactions: a booster shot for the cosmic neutrino background}

\author{Anirban Das\,\orcidlink{0000-0002-7880-9454}}
\email{anirband@slac.stanford.edu}
\affiliation{SLAC National Accelerator Laboratory, Stanford University, 2575 Sand Hill Road, Menlo Park, CA 94025, USA}
\author{Yuber F. Perez-Gonzalez\,\orcidlink{0000-0002-2020-7223}}
\email{yuber.f.perez-gonzalez@durham.ac.uk}
\affiliation{Institute for Particle Physics Phenomenology, Durham University, South Road DH13EL, Durham, United Kingdom}

\author{Manibrata Sen\,\orcidlink{0000-0001-7948-4332}}
\email{manibrata@mpi-hd.mpg.de}
\affiliation{Max-Planck-Institut f\"ur Kernphysik, Saupfercheckweg 1, 69117 Heidelberg, Germany}

\preprint{SLAC-PUB-17673, IPPP/22/24}
\begin{abstract}
    Neutrinos might interact among themselves through forces that have so far remained hidden.
    Throughout the history of the Universe, such \emph{secret} interactions could lead to scatterings between the neutrinos from supernova explosions and the non-relativistic relic neutrinos left over from the Big Bang. Such scatterings can boost the cosmic neutrino background (\cnub) to  energies of  ${\cal O}$(MeV), making it, in principle, observable in experiments searching for the diffuse supernova neutrino background.
    Assuming a model-independent\COM{, but flavor universal,} four-Fermi interaction, we determine the upscattered cosmic neutrino flux, and derive constraints on such secret interactions from the latest results from Super-Kamiokande. Furthermore, we also study prospects for detection of the boosted flux in future lead-based coherent elastic neutrino-nucleus scattering experiments. 
    \COM{Nevertheless, given current constraints on flavor universal self-interactions, we find that the upscattered \cnub~contribution to the total DSNB flux is negligible, making a possible measurement of the boosted C$\nu$B insurmountable.}
\end{abstract}

\maketitle

\section{Introduction}
The Standard Model of particle physics has come a long way since Pauli's proposal of the elusive neutrino in 1930. Not only have neutrinos been detected, but also it has been proven conclusively that these elementary particles have a tiny mass, thereby allowing different flavors of neutrinos to oscillate among each other~\cite{ParticleDataGroup:2018ovx}. While three flavors of neutrinos have been shown to exist, tremendous amount of theoretical and experimental efforts have been underway to explore the existence of additional families of neutrinos. Neutrinos are known to interact with other elementary particles as well as themselves through weak interactions. However, what is not known conclusively is whether there exists secret interactions among neutrinos.

The hypothesis for such secret self-interactions of neutrinos has a plethora of consequences for the early Universe, astrophysical setups like compact objects as well as terrestrial experiments (see Ref.\,\cite{Berryman:2022hds} for a comprehensive recent review).
Laboratory bounds on such self-interactions can arise from different sources - Higgs invisible decay bounds, constraints arising from Z boson decay width, decays of $\pi,K,D$ mesons, as well as neutrinoless double beta decay searches~\cite{Brune:2018sab,Berryman:2018ogk,Blinov:2019gcj,DeGouvea:2019wpf,Deppisch:2020sqh,Brdar:2020nbj,Kelly:2020pcy}. 
From a cosmological perspective, a strong bound on these self-interactions exists from the successful predictions of primordial abundances of elements during big-bang nucleosynthesis~\cite{Huang:2017egl,Grohs:2020xxd}, as well as from its impact on the CMB~\cite{Bell:2005dr,Archidiacono:2013dua,Cyr-Racine:2013jua,Oldengott:2014qra,Oldengott:2017fhy,Lancaster:2017ksf,Forastieri:2019cuf,Kreisch:2019yzn,Das:2020xke,Brinckmann:2020bcn,RoyChoudhury:2020dmd,Das:2021guu,Venzor:2022hql} and structure-formation~\cite{Hannestad:2004qu}. Astrophysical probes of such strong self-interactions can arise from stellar cooling arguments~\cite{Escudero:2019gzq}, core-collapse supernovae (CCSN)~\cite{Kolb:1987qy,Kachelriess:2000qc,Farzan:2002wx,Farzan:2014gza,Das:2017iuj,Dighe:2017sur,Shalgar:2019rqe}, as well as the observation of high-energy neutrinos in neutrino telescopes~\cite{Hooper:2007jr,Ioka:2014kca,Kelly:2018tyg,Murase:2019xqi,Barenboim:2019tux,Bustamante:2020mep,Esteban:2021tub}. The introduction of secret self-interactions of active neutrinos can affect the dynamics of the core-collapse, leading to a delayed emission of neutrinos~\cite{Fuller:1988ega,Chang:2022aas}. From a flavor point of view, it is important to appreciate that the most stringent bounds exist in the electron and the muon sector, with substantially weaker bounds for tau-neutrinos. Self-interactions of active neutrinos have also been shown to assist in successful production of sterile neutrino dark matter, alleviating astrophysical bounds arising from un-observation of X-rays~\cite{DeGouvea:2019wpf,Kelly:2020pcy,Kelly:2020aks, Alonso-Alvarez:2021pgy,Sen:2021mxl,Benso:2021hhh}. 

One particular direction remains unexplored, nevertheless. The Universe is filled with an abundance of thermalized relic neutrinos, which decoupled when the Universe had a temperature of $\sim 1\MeV$, and have been free-streaming since then. This cosmic neutrino background (\cnub{}) consists of non-relativistic neutrinos\footnote{If the lightest neutrino has a mass $m_0\lesssim 0.6$~meV, it would still be relativistic today.}, with a temperature of around $\mathcal{O}(0.1)$~meV. On the other hand, the Earth also receives a considerable flux of relativistic neutrinos from different sources, for example, an isotropic sea of MeV neutrinos from the diffuse supernova neutrino background (DSNB), higher energy neutrinos from cosmogenic sources, and so on. In the presence of large secret self-interactions, these relativistic neutrinos can scatter with the \cnub{}, and boost them to higher energies. This can provide a healthy fraction of \emph{relativistic} \cnub~ flux, with energies comparable to those of the scattering neutrinos, and hence can be detected in current as well as future neutrino detectors.

In this work, we demonstrate this idea with the help of the DSNB-\cnub{} scattering. As mentioned before, the DSNB consists of MeV neutrinos of all flavors emanating from all CCSNe since the beginning of the star-formation epoch\,\cite{Beacom:2010kk}. After interacting, the neutrinos from the \cnub{}
get scattered to MeV energies, and thus can be probed by experiments sensitive to the DSNB. The boosted \cnub{} spectra will add on to the DSNB spectra, and can lead to an increased number of events detected at lower energies. On the experimental front, a lot of effort has been underway towards the detection of the DSNB. Super-Kamiokande (SK)~\cite{Super-Kamiokande:2021jaq} in Japan, loaded with Gadolinium, leads the pack, followed by upcoming experiments like the Jiangmen Underground Neutrino Observatory (JUNO) in China~\cite{JUNO:2015zny}, Deep Underground Neutrino Obervatory (DUNE) in the USA~\cite{DUNE:2020ypp}, Hyper-Kamiokande (HK) --- an upgrade of the SK detector~\cite{Hyper-Kamiokande:2018ofw}, among others. A number of other ideas like water-based liquid scintillator detectors~\cite{Askins:2019oqj,Gann:2015fba}, paleo-detectors~\cite{Baum:2022wfc}, as well as using archaeological lead~\cite{Pattavina:2020cqc} for detection are being explored. The possibility of using coherent elastic neutrino-nucleus scattering to detect the DSNB has also been explored recently~\cite{Suliga:2021hek}. We utilize some of these experiments to probe the boosted \cnub{} spectra, and use the results to put constraints on secret neutrino self-interactions. A similar study about the resonant scattering between the high energy and cosmic neutrinos was done in Ref.\,\cite{Creque-Sarbinowski:2020qhz}.

The paper is organized as follows. In Section \ref{sec:NSelf}, we introduce neutrino self-interactions and establish the notation for the following sections.
In section \ref{sec:DSNB-CnB}, we describe briefly the standard properties of the DSNB and the \cnub.
We then describe the computation of the upscattered \cnub{} flux in section \ref{sec:BCnuB}.
Using the computed boosted flux, in section \ref{sec:resl}, we determine the current constraints from SK, and analyse the future prospects for detection in a lead-based coherent elastic neutrino-nucleus scattering experiment.
Finally, we draw our conclusions in section \ref{sec:conc}.
We use natural units in which $\hbar = c = k_{\rm B} = 1$ throughout this manuscript.
\section{Neutrino self-interactions}\label{sec:NSelf}
Neutrino self-interactions can arise due to exchange of a new scalar, and/or a vector particle. Scalar mediated self-interactions can arise from the following higher-dimensional operator involving the Higgs doublet $(H)$ and the lepton doublet $(L)$~\cite{Berryman:2018ogk},
\begin{equation}
    \mathcal{L}\supset \frac{f^\varphi_{\alpha\beta}}{\Lambda^2}\left(\overline{L}_\alpha\, H\right)\left(H^T\, L^C_\beta\right) \varphi^*,\,
    \label{eq:LagrangianScalEFT}
\end{equation}
where $\varphi$ is a complex scalar with nonzero lepton-number to avoid violating such a symmetry. After the electroweak symmetry breaking, one arrives at the following neutrinophilic interaction,
\begin{equation}
    \mathcal{L}\supset g^{\varphi}_{\alpha\beta}\,\overline{\nu}_\alpha\nu^C_\beta \varphi^*,\,
    \label{eq:Lagrangian}
\end{equation}
where $g^{\varphi}_{\alpha\beta}=f^{\varphi}_{\alpha\beta} v^2/\Lambda^2$. 

Similarly, for vector-boson mediated self-interactions, we consider a gauge boson $V$ which couples only to the active neutrinos at low energies through the following operator~\cite{Kelly:2020pcy},
\begin{equation}
    \mathcal{L}\supset \frac{f^{\rm V}_{\alpha\beta}}{\Lambda^2}\left(\overline{L}_\alpha\,i\sigma_2 H^*\right)\gamma^\mu \left(H^T\, i\sigma_2 L_\beta\right) V_\mu\,,
    \label{eq:LagrangianVecEFT}
\end{equation}
where the cut-off scale $\Lambda$ signifies the  energy scale close to which the effective theory breaks down. Once the Higgs attains a vacuum expectation value, the above operator leads to active neutrino self-interactions of the form,
\begin{equation}
    \mathcal{L}\supset g^{\rm V}_{\alpha\beta}\,\overline{\nu}_\alpha\gamma^\mu\nu_\beta V_\mu\,,
    \label{eq:Lagrangian}
\end{equation}
where $g^{\rm V}_{\alpha\beta}=f^{\rm V}_{\alpha\beta}v^2/\Lambda^2$. 

In this paper, we shall mostly be concerned with scattering energy ($\lesssim 10\MeV$) below the mediator mass. In that case, the neutrino self-interaction can be described by four-Fermi interaction and the scattering cross sections can always be approximated as \begin{align}\label{eq:cross_section}
    \sigma \approx \frac{(g_{\alpha\beta}^{\varphi,V})^4 s}{16\pi M_{\varphi,V}^4}\,\equiv G_X^2m_\nu E_{\nu}\,.
\end{align}
Here $s=2m_\nu E_\nu$ is the center-of-mass energy when the neutrino from the DSNB has energy $E_\nu$ and the \cnub{} is almost at rest, and the effective coupling $G_X=\sqrt{2/\pi}(g_{\alpha\beta}^{\varphi,V})^2/(4M_{\varphi,V}^2)$ being determined by the nature of the interaction. 
This $E_\nu$-scaling in Eq.(\ref{eq:cross_section}) is inspired by the exact expression of a four-Fermi interaction cross section when one particle is at rest.

\begin{figure}[!t]
\includegraphics[width=0.45\textwidth]{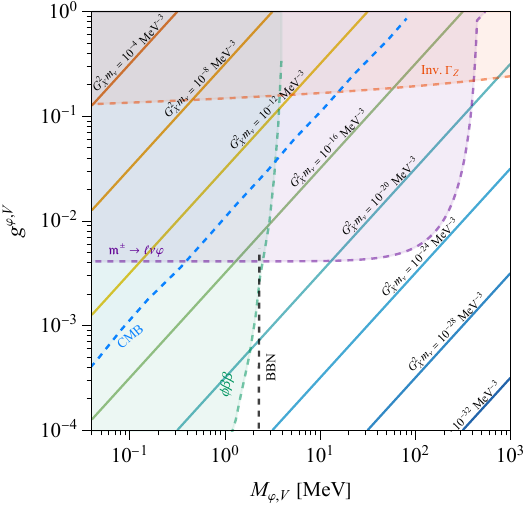}
    \caption{Isocontours of $G_X^2m_\nu$ for the scalar case in the plane $g_{\alpha\beta}^{\varphi}$ vs $M_{\varphi}$. We present constraints from Invisible $Z$ decay (dark orange region), meson decays $\mathfrak{m}^\pm\to \ell \nu\varphi$ (purple), and double-beta decay (green), taken from \cite{Berryman:2018ogk,Berryman:2022hds}. The cosmological constraints coming from CMB and BBN measurements have been taken from~\cite{Sabti:2019mhn}.}
\label{fig:cont}
\end{figure}

We also focus on flavor universal couplings such that $g_{ee}=g_{\mu\mu}=g_{\tau\tau}$, while the flavor off-diagonal couplings are set to zero. Note that in such a case, the couplings are identical in the flavor and the mass bases\footnote{The distinction between flavor and mass basis is crucial since at the time of scattering, the neutrinos from the DSNB and well as the \cnub{} have decohered and behave as pure mass-eigenstates}. In Fig.\,\ref{fig:cont}, we show the contours of a few values of $G_X^2m_\nu$, that we shall be using in this paper, in the plane of $g^{\varphi}-M_{\varphi}$ for the scalar mediator. The results for the vector mediators are similar, see~\cite{Kelly:2020pcy} for further details. We assume $m_\nu=1\eV$ for illustration purpose here.
Note that such a value is currently disfavored by CMB data\,\cite{Aghanim:2018eyx}, and is also slightly above the current KATRIN bounds on the effective neutrino mass~\cite{KATRIN:2021uub}. Other experimental constraints are also shown for reference.
We stress that modifying the flavor structure of the interactions will modify the constraints presented in Fig.~\ref{fig:cont}. For instance, allowing for interactions only with tau neutrinos removes the $\mathfrak{m}^\pm\to \ell \nu\varphi$ constraint. While this is certainly a new direction to explore, it merits a dedicated analysis of its own. Hence, we do not want to go into its details in this work.
Heretofore, we stay agnostic of the nature of the mediator when computing the boosted \cnub{} spectra. The formalism presented below is generic, and can be used for either case. We will specify the nature of the mediator only when translating experimental probes to constraints on the mediator mass-coupling parameter space. We also will consider values of $G_X^2m_\nu$ that might be currently excluded in order to determine the capabilities of DSNB searches to constrain self-interactions.
\section{DSNB and \cnub}\label{sec:DSNB-CnB}
Each CCSN emits a near-thermal flux of neutrinos that can be approximated by a Fermi-Dirac distribution with a flavor-dependent temperature $T_\nu$\,\cite{Beacom:2010kk},
\begin{equation}
    F_\nu(E)=\frac{E_\nu^\mathrm{tot}}{6}\,\frac{120}{7\pi^4}\,\frac{E_\nu^2}{T_\nu^4}\,\frac{1}{e^{E_\nu/T_\nu}+1}\,,
\end{equation}
\begin{center}
 \begin{table}[!t]
 \centering
 \caption{The star formation rate parameters used in Eq.(\ref{eq:sfr}). The uncertainties are taken from Ref.\,\cite{Horiuchi:2008jz}.}\label{tab:sfr_params}
 \vspace{0.2cm}
 \begin{tabular}{|C{2.15cm}|C{5cm}|}
 \hline\hline
 Parameter & Value\\ \hline
 &\\[-1.5ex]
 $\rho_*$ & $0.02_{-0.006}^{+0.001}\ms \mathrm{~yr^{-1}Mpc^{-3}}$\\[1ex]
 $a$ & $3.4\pm0.2$\\[1ex]
 $b$ & $-0.3\pm0.12$\\[1ex]
 $c$ & $-2.5\pm1$\\[1ex]
 $z_1$ & $1$\\[1ex]
 $z_2$ & $4$\\[1ex]
 $\zeta$ & $-10$\\[1ex]
 \hline
 \end{tabular}
\end{table}
\end{center}
where $E_\nu^\mathrm{tot}$ is the total energy in all neutrinos. We assume $T_{\nu_e}=6.6\MeV, T_{\bar{\nu}_e}=7\MeV,$ and $T_{\nu_x}=10\MeV$ unless otherwise stated. We will explore the effect of having different temperatures when determining current constrains from SK. The net DSNB flux received at earth can be obtained by integrating the fluxes from each individual SN that are distributed over a range of redshifts. The redshift distribution of the SNe is proportional to the star formation rate which can be fitted with a broken power law\,\cite{Kistler:2013jza},
\begin{equation}\label{eq:sfr}
\begin{split}
    R_\mathrm{CCSN}(z) &= \frac{\rho_*}{143\ms}\\
    &\times\left((1+z)^{a\zeta}+\left(\frac{1+z}{B}\right)^{b\zeta}+\left(\frac{1+z}{C}\right)^{c\zeta}\right)^{1/\zeta}\,,
\end{split}
\end{equation}
where $B=(1+z_1)^{1-a/b}, C=(1+z_1)^{(b-a)/c}(1-z_2)^{1-b/c}$. The numerical values of the various parameters are given in Table\,\ref{tab:sfr_params}. The DSNB flux at a redshift $z$ can be obtained by convolving the CCSN rate with the neutrino flux from an individual SN including the expansion of the Universe,
\begin{equation}
    \Phi(E_\nu,z) = (1+z)^2 \int_z^\zmax \frac{dz'}{H(z')} F_\nu[E_\nu(1+z')] R_\mathrm{CCSN}(z')\,.
\end{equation}
We multiply $E_\nu$ by an appropriate factor of $(1+z')$ as the the neutrinos were emitted at that redshift. The Hubble parameter is given by $H(z) = H_0 \sqrt{\Omega_\Lambda+\Omega_M(1+z)^3}$ with $H_0=67\,\mathrm{km~s^{-1}Mpc^{-1}}$ as per the Planck 2018 data\,\cite{Aghanim:2018eyx}. We integrate the SN formation rate up to redshift $\zmax=6$. As an illustration, in Fig.\,\ref{fig:CnBflux}, we show the DSNB spectrum for $\overline{\nu}_e$ at $z=0$. 

The \cnub{} is the population of relic neutrinos that were frozen out after decoupling from the thermal plasma. The weak interactions of the SM neutrinos became slower than the Hubble expansion rate around a photon temperature $T_\gamma \simeq 1\MeV$. Since then, neutrinos have been free-streaming and cooling down as a result of the expansion of the Universe. They are believed to have a Fermi-Dirac distribution today with a temperature $T_\nu \simeq 1.95\K$. Hence the number density of \cnub{} is given by
\begin{equation}
    n_\nu = \frac{3}{4}\frac{\zeta(3)}{\pi^2}gT_\nu^3\,,
\end{equation}
where $g$ is the number of degrees of freedom. Looking at their temperature today, it is clear that these neutrinos have very low energy ($T_\nu \simeq 1.95\K \approx 0.17\,\mathrm{meV}$) compared to the DSNB. Their temperature $T_\nu$ today is slightly lower than the photons as the electrons and positrons annihilated after neutrino decoupling and deposited their energy into the photon bath heating it up approximately by a factor of $(11/4)^{1/3}$. Although we do not know the absolute masses of the SM neutrinos, the atmospheric mass difference $\Delta m_\mathrm{atm}^2=2.5\times 10^{-3}\,\mathrm{eV^2}$ tells us that at least two neutrino mass states are non\-re\-la\-ti\-vis\-tic today\,\cite{Esteban:2020cvm}.
\section{Boosted flux}\label{sec:BCnuB}
As discussed before, in the presence of neutrino self-interactions, the supernova relic neutrinos scatter off the \cnub, and transfer their energy to the latter. This allows a fraction of the non-relativistic \cnub{} to get boosted to higher energies, comparable to those of the DSNB. This can have interesting consequences for experiments searching for the DSNB. 

Neutrinos originating from a SN at a redshift $z'\in (0-\zmax)$ will scatter off the \cnub{} at some later redshift $z''\leq z'$. The scattered DSNB flux as a function of energy $E_D$ at redshift $z$ can then be estimated as 
\begin{widetext}
\begin{equation}
	\frac{d\Phi_D(z)}{dE_D} = (1+z)^2\int_z^\zmax \frac{dz'}{H(z')}  R_\mathrm{CCSN}(z')  F_\nu[E_D(1+z')]
	\times\exp\left[{-\int_{z}^{z'} \frac{dz''}{(1+z'')H(z'')} n_\nu(z'') \sigma}\right]\,,
	\label{eq:scattDSNB}
\end{equation}
\end{widetext}
where $n_\nu(z'')$ is the number density of the \cnub{} at the redshift $z''$. At the same time, the non-relativisitic neutrinos from the \cnub{} gets upscattered to an energy $E_C$. Because the neutrinos from the DSNB carry $\mathcal{O}(10\MeV)$ energy, the maximum energy transferred to \cnub, ignoring the mass of neutrino, is simply $E_C^\mathrm{max}\approx E_D$. The upscattered \cnub{} flux can be computed as 
\begin{equation}
	\frac{d\psi_C}{dE_C} = \int_{E_C}^\infty  dE_D \int_0^\zmax \frac{dz}{H(z)} n_\nu(z)\,\frac{d\sigma}{dE_C}\, \frac{d\Phi_D(z)}{dE_D},
	\label{eq:scattCNB}
\end{equation}
In the heavy mediator limit, i.e., when the mediator mass is heavier than the energy transferred, the differential cross section can be simplified to $d\sigma/dE_C = \sigma/E_C^\mathrm{max}$. Eq.\,\ref{eq:scattCNB} will be used to estimate the upscattered \cnub{} flux. The details of modelling the self-interactions enter through the cross-section $\sigma$. 
\begin{figure}[!t]
\includegraphics[width=0.425\textwidth]{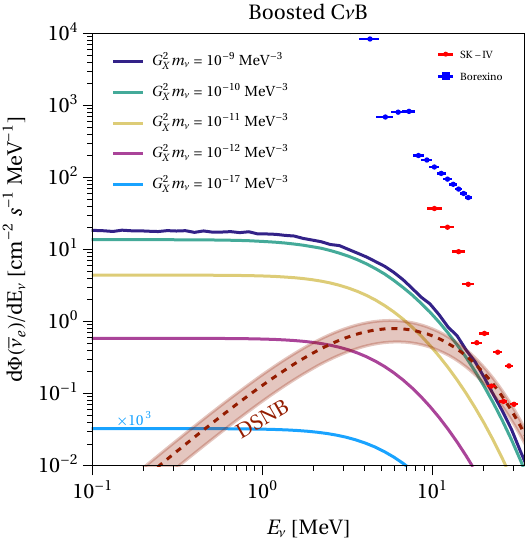}
    \caption{$\overline{\nu}_e$ flux as function of neutrino energy for the standard DSNB (dark orange band) and the upscattered \cnub flux for four different values of $G^2_X m_\nu=10^{-9}~{\rm MeV^{-3}}$ (purple), $10^{-10}~{\rm MeV^{-3}}$ (emerald), $10^{-11}~{\rm MeV^{-3}}$ (dark yellow), $10^{-12}~{\rm MeV^{-3}}$ (lilac)\COM{, and $10^{-17}~{\rm MeV^{-3}}$ (light blue). Note that we have multiplied the last flux by a factor of $10^3$ to make it visible.} We present current constrains on $\overline{\nu}_e$ searches from Borexino~\cite{Borexino:2019wln} (blue) and SK-IV~\cite{Super-Kamiokande:2021jaq} (red).}
\label{fig:CnBflux}
\end{figure}

Fig.~\ref{fig:CnBflux} depicts the upscattered $\bar{\nu}_e$ component of the \cnub{} for different values of $G^2_X m_\nu$. The corresponding $\bar{\nu}_e$ from the DSNB is also shown for comparison. Note that, for clarity, we have only plotted the upscatttered \cnub{} flux for the central values of the SFR parameters in Table\,\ref{tab:sfr_params}. We find that for $G^2_X m_\nu>10^{-9}\,{\rm MeV}^{-3}$, the boosted cosmic neutrino flux dominates over the SN relic neutrino flux for energies $E_{\nu}>1\MeV$. Since $E_D\gg E_C$ and the scattering particles have the same mass, almost all of the SN relic neutrino energy gets transferred to the cosmic neutrino flux; hence the upscattered flux is independent of energy for low energies, and follows the DSNB spectra for $E_\nu>$ few MeVs. The corresponding limits from searches of $\bar{\nu}_e$ in these energy ranges from SK~\cite{Super-Kamiokande:2021jaq} and Borexino~\cite{Borexino:2019wln} are also shown. While constraints from Borexino are weaker, and can be probed with larger values of $G^2_X m_\nu$, the current bounds from SK can already probe such a boosted $\bar{\nu}_e$ flux. However, as can be seen from Fig.~\ref{fig:cont}, such large flavor-universal interaction strength is already excluded by laboratory constraints from $Z$ and meson decays.
\COM{For strength values still allowed, $G^2_X m_\nu \lesssim 10^{-17}\,{\rm MeV}^{-3}$, see Fig.~\ref{fig:cont}, the boosted \cnub{} flux becomes negligible; the flux for $G^2_X m_\nu = 10^{-17}\,{\rm MeV}^{-3}$ in Fig.~\ref{fig:CnBflux} has been multiplied by a factor of $10^3$ to make it visible the figure.
}

\section{Detection and limits}\label{sec:resl}

\begin{figure}[!t]
\includegraphics[width=0.425\textwidth]{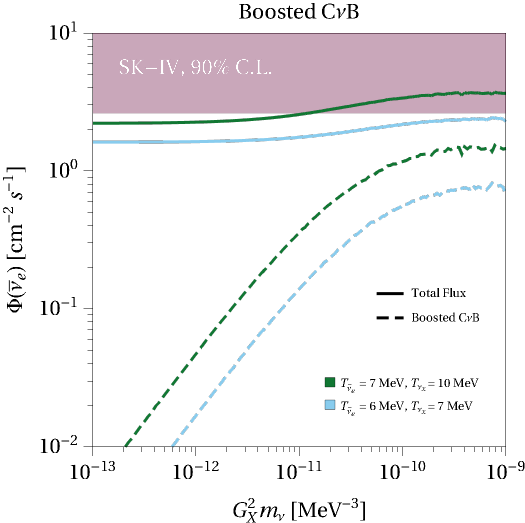}
\caption{Total $\overline{\nu}_e$ flux for $E_\nu>17.8$~MeV as function of $G^2_X m_\nu$ for two different sets of SN temperatures, $T_{\bar{\nu}_e}=7\MeV,$ and $T_{\nu_x}=10\MeV$ (green) and $T_{\bar{\nu}_e}=6\MeV,$ and $T_{\nu_x}=7\MeV$ (light blue). 
The full lines correspond to the total DSNB+boosted \cnub{} fluxes, while the dashed curves depict the contribution from the upscattered cosmic neutrinos.
The dark lilac band corresponds to the SK-IV limit at the 90\% C.L.}
\label{fig:SKlimit}
\end{figure}
Fig.~\ref{fig:SKlimit} shows the current limit of 90\% C.L. from SK-IV searches of the DSNB as function of $G^2_X m_\nu$. SK will be sensitive to the net neutrino flux (solid lines) consisting of the original DSNB flux, as well as the boosted \cnub{} flux (dashed lines). The green and the blue lines correspond two different choices of $T_{\bar{\nu}_e}=7\MeV,\, T_{\nu_{\mu,\tau}}=10\MeV$, and $T_{\bar{\nu}_e}=6\MeV,\, T_{\nu_{\mu,\tau}}=7\MeV$. We find that the former case, characterized by slightly larger temperatures than what is expected from SN simulations, is already ruled out by SK-IV for $G^2_X m_\nu\gtrsim 10^{-11}{\rm MeV}^{-3}$. This can be directly translated to bounds on the mediator mass and couplings for a specific model of neutrino self-interactions. The case corresponding to lower values of $(T_{\bar{\nu}_e},T_{\nu_{\mu,\tau}})$ stays out of reach of the current SK bounds, but can be probed by future surveys.
Indeed, SK intends to constrain $\Phi_{\overline{\nu}_e}\lesssim 1~{\rm cm^{-2} s^{-1}}$ for $E_\nu\gtrsim 17.3$~MeV in the next ten-years, after doping their detector with Gadolinium~\cite{Super-Kamiokande:2021jaq}.
Furthermore, we find that there is little sensitivity to values of $G^2_X m_\nu< 10^{-11}{\rm MeV}^{-3}$. In this case, the boosted component of the \cnub{} flux is too tiny to make any appreciable difference to the DSNB flux.

\begin{figure}[!t]
\includegraphics[width=0.45\textwidth]{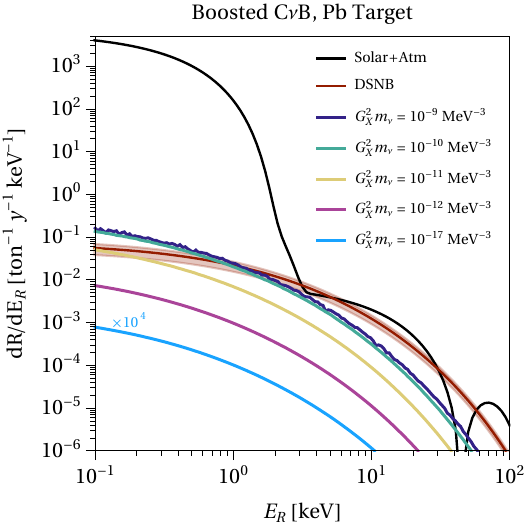}
    \caption{Expected CEvNS differential recoil rate for a Pb target as function of the nuclear recoil energy.
    We present the recoil rates produced by the upscattered \cnub{} flux for four different values of $G^2_X m_\nu=10^{-9}~{\rm MeV^{-3}}$ (purple), $10^{-10}~{\rm MeV^{-3}}$ (emerald), $10^{-11}~{\rm MeV^{-3}}$ (dark yellow), $10^{-12}~{\rm MeV^{-3}}$ (lilac)\COM{, and $10^{-17}~{\rm MeV^{-3}}$ (light blue). Note that we have multiplied the last case by a factor of $10^4$}.
    The black line represents the rate for solar + atmospheric neutrinos, while the dark orange band corresponds to the prediction for the DSNB.}
\label{fig:CevNS}
\end{figure}

\begin{figure}[!t]
\includegraphics[width=0.45\textwidth]{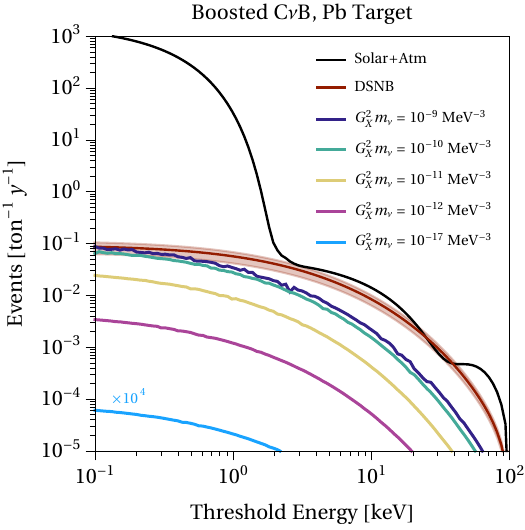}
\caption{Expected CEvNS number of events for a Pb      target as function of the experimental threshold energy, produced by the upscattered \cnub{}  flux for four different values of $G^2_X m_\nu=10^{-9}~{\rm MeV^{-3}}$ (purple), $10^{-10}~{\rm MeV^{-3}}$ (emerald), $10^{-11}~{\rm MeV^{-3}}$ (dark yellow), $10^{-12}~{\rm MeV^{-3}}$ (lilac)\COM{, and $10^{-17}~{\rm MeV^{-3}}$ (light blue). Note that we have multiplied the last case by a factor of $10^4$}.
The black line represents the events for solar + atmospheric neutrinos, while the dark orange band corresponds to the prediction for the DSNB.}
\label{fig:CevNS_Evs}
\end{figure}
\begin{figure*}[!t]
\includegraphics[width=0.85\textwidth]{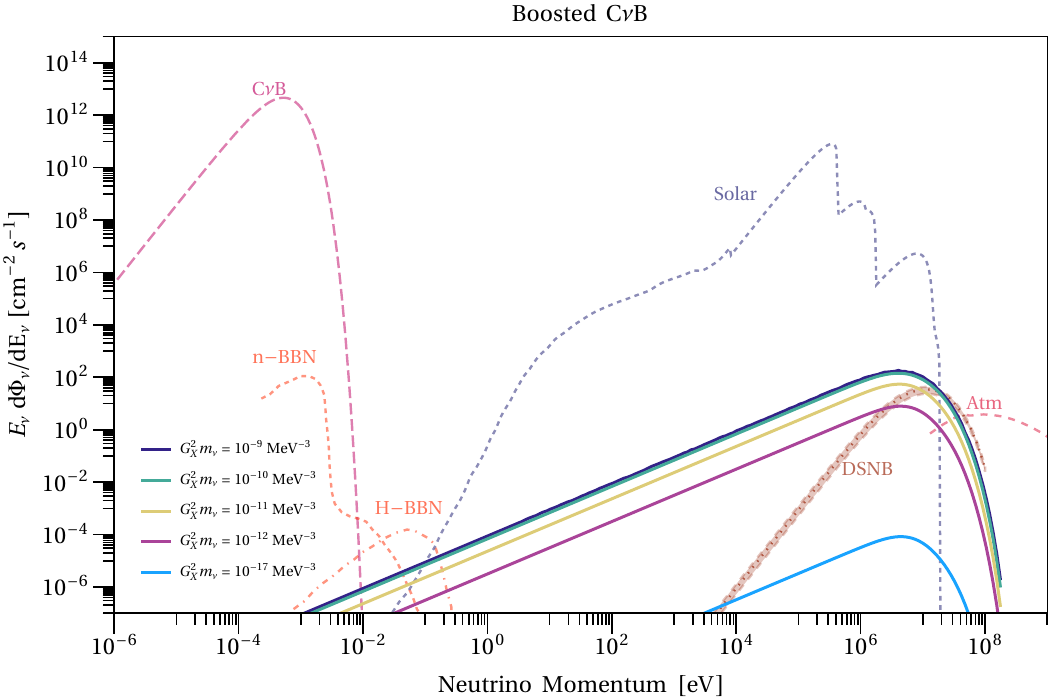}
\caption{Total neutrino flux times energy as function of neutrino momentum for the upscattered \cnub{} flux for four different values of $G^2_X m_\nu=10^{-9}~{\rm MeV^{-3}}$ (purple), $10^{-10}~{\rm MeV^{-3}}$ (emerald), $10^{-11}~{\rm MeV^{-3}}$ (dark yellow), $10^{-12}~{\rm MeV^{-3}}$ (lilac)\COM{, and $10^{-17}~{\rm MeV^{-3}}$ (light blue)}. 
For comparison, we present the standard \cnub{} (dashed lilac), relic neutrinos from BBN (orange dashed and dot-dashed), solar neutrinos (blue dashed), and low-energy atmospheric neutrinos (red dashed). These neutrino fluxes are obtained from~\cite{Vitagliano:2019yzm}.}
\label{fig:FullFlux}
\end{figure*}

Recent works have demonstrated the feasibility of observing the DSNB via the Coherent Elastic Neutrino-Nucleus Scattering (CEvNS) in future
experiments~\cite{Suliga:2021hek,Baum:2022wfc}.
Specifically, the RES-NOVA proposal will exploit the enhancement of the CEvNS cross section for a heavy nucleus like lead, together with the high purity of archaeological Pb~\cite{Pattavina:2020cqc}, in such a way that it will be possible to search for the DSNB.
Therefore, we can expect that such facilities could also search for the boosted \cnub{} flux from our scenario.
We estimate the recoil rates and number of events in a generic experiment using Lead as follows.
The recoil rate is computed via
\begin{align}
    \frac{dR}{dE_R}={\cal E}\int dE_C \frac{d\sigma_{\rm CEvNS}}{dE_R}\frac{d\psi_C}{dE_C}\,,
\end{align}
where ${\cal E}$ is the total experimental exposure (observation time $\times$ number of targets), $E_R$ is the nuclear recoil energy, and $d\sigma_{\rm CEvNS}/dE_R$ is the CEvNS cross section, which for completeness we quote next,
\begin{align}
    \frac{d\sigma_{\rm CEvNS}}{dE_R} = \frac{G_F^2 m_X}{4\pi} {\cal Q}_W^2 \left(1-\frac{m_XE_R}{2E_C^2}\right){\cal F}(Q^2),
\end{align}
being $G_F$ the Fermi constant, $m_X$ the nuclear target mass, $ {\cal Q}_W = N-Z(1-4\sin^2\theta_W)$ the weak vector nuclear charge, with $N,Z$ the number of neutrons, nucleons, respectively, and ${\cal F}(Q^2)$ the nuclear form factor, taken here to be the Helm form factor~\cite{PhysRev.104.1466}.
The number of CEvNS events from the upscattered \cnub{} flux is therefore obtained by integrating the recoil rate in the recoil energy,
\begin{align}
    {\cal N}_C=\int_{E_{\rm th}} dE_R \frac{dR}{dE_R}\,,
\end{align}
where the integration is performed from the experimental energy threshold $E_{\rm th}$.

We present in Fig.\,\ref{fig:CevNS} the recoil rate and number of events for the boosted \cnub{} flux for four different values of $G^2_X m_\nu$ as function of the recoil and threshold energies, respectively.
We also show the total solar+atmospheric neutrinos recoil rate (black) and the DSNB contribution (purple band).
For the recoil rate, we observe that the upscattered \cnub{} contribution is above the standard CEvNS from the DSNB at very low recoil energies ($E_R\lesssim 1$ keV) for $G^2_X m_\nu\gtrsim 10^{-10}~{\rm MeV^{-3}}$. 
Nevertheless, the enormous contribution coming from solar neutrinos makes the search in such ranges of recoil energies rather difficult. 
For larger recoil energies, we find that the recoil events from the DSNB become larger than the boosted contribution.
However, in the region $E_R\in [2-3]$~keV the DSNB rate is of the same order as the atmospheric neutrino rate.
Meanwhile, for $G^2_X m_\nu\gtrsim 10^{-9}~{\rm MeV^{-3}}$, the \cnub{} rate is only a factor of $\sim 70\%$ the DSNB value.
Thus, in principle, detailed searches in this energy range could aid in exploring the parameter space of secret self-interactions.

Such a behavior is reflected in the number of events dependent on the energy threshold, Fig.~\ref{fig:CevNS_Evs}.
For relatively large energy thresholds, above the solar neutrinos, we observe that the DSNB contribution is about half the atmospheric one.
Indeed, for a threshold of $E_{\rm th}\sim 3$ keV, larger than the expected 1 keV threshold in the RES-NOVA proposal~\cite{Pattavina:2020cqc}, the events for the fiducial value of the DSNB would be $\sim 0.015$ per ton$\cdot$y while the atmospheric background corresponds to $\sim 0.03$ per ton$\cdot$y. 
Meanwhile, the events from the boosted \cnub{} neutrinos would be $\sim 0.007$ per ton$\cdot$y.
We can estimate that at least an exposure of ${\cal E}\sim 1$~kton$\cdot$y would be necessary to have $S/\sqrt{B}\sim 1$ for $G^2_X m_\nu\gtrsim 10^{-9}~{\rm MeV^{-3}}$, including the standard DSNB contribution as background.
For smaller $G^2_X m_\nu$, we would require much larger exposures, ${\cal E}\sim 20$~kton$\cdot$y, for $G^2_X m_\nu\gtrsim 10^{-11}~{\rm MeV^{-3}}$.
Surely, these are na\"ive estimates; for much more realistic estimates, we would require introducing additional backgrounds and other experimental characteristics of RES-NOVA.
We leave such an analysis for future work.

From the preceding discussion, we infer that significant boosting of cosmic neutrino background is possible only for relatively large interaction strength $G^2_X m_\nu\gtrsim 10^{-9}~{\rm MeV^{-3}}$. However, such couplings are in tension with various laboratory experiments as discussed before. Hence the boosted cosmic neutrinos, with allowed interaction strength, would only make up a subdominant part of the MeV-scale neutrino flux.


%
\section{Conclusions}\label{sec:conc}
Currently, only neutrinos hold the guaranteed key to the gateway of physics beyond the Standard Model. Search for secret neutrino self-interaction is a promising avenue that might lead us to discovery of such BSM physics.
Current laboratory bounds on these interactions allow for the possibility of large couplings, which can even be stronger than the weak interaction by several orders of magnitude. This presents a perfect opportunity to study the effect that strongly self-interacting neutrinos can have in different areas.

In this paper, we study a simple yet important outcome of strongly self-interacting neutrinos. The Universe is filled with a thermal bath of non-relativistic neutrinos (\cnub), which have decoupled when the Universe was quite hot, at a temperature of around $1$~MeV. Another ubiquitous source of MeV-energy (relativistic) neutrinos is all the core-collapse supernovae that have happened in the past.
In the presence of large non-standard self-interactions, the DSNB will scatter with the \cnub, and upscatter the latter to relativistic energies. We compute this upscattered flux. The results, shown in a larger picture in Fig.\,\ref{fig:FullFlux} for five different cases, demonstrate how the upscattered \cnub{} flux compare to the original one, as well as the fluxes from other sources. The upscattered flux is comparable to, and in some cases even larger than, the DSNB flux, however, for the strength of the self-interaction required for such upscattering is already ruled out by current experiments. 
\COM{For the values still allowed by current data, $G^2_X m_\nu=10^{-17}~{\rm MeV^{-3}}$, we observe that the resulting upscattered \cnub{} flux is orders of magnitude below the DSNB and solar neutrinos, which makes a discovery rather impractical.
Let us stress, however, that we have assumed flavor universal couplings. If the self-interactions have a different flavor dependence, some of the laboratory constraints might not be applicable, and the boosted flux could be larger.
Thus,} using the most recent search of the DSNB by the Super-Kamiokande collaboration $(22.5\times 5823)$~kton$\cdot$days of data, we put model-independent constraints on the combination of the coupling, and the mass of the mediator which mediates the self-interactions.
We emphasize that this constitutes a direct test of the impact of the secret neutrino self-interactions. 

Furthermore, we also computed the expected event rate for the net boosted \cnub{} using coherent scattering on a generic lead-based detector. We find that for a reasonable choice of parameters, the upscattered \cnub{} flux can be comparable to the DSNB flux at lower recoil energies, while for larger energies the DSNB flux dominates.
A na\"ive estimate indicates that an exposure of $\sim 400$~ton$\cdot$y, such as the expected by the RES-NOVA collaboration in the future stage-3 of the project~\cite{Pattavina:2020cqc}, would be needed to start exploring the parameter space in our scenario. However, obtaining competitive constraints in comparison to current limits seems rather difficult to achieve with this technology.
For smaller values of the mediator mass (which would be in tension with BBN in standard cosmologies), one can get stronger self-interaction cross-sections. In such a scenario, the boosted \cnub{} flux can dominate over the DSNB, and hence already by probed by experiments. However, theoretically it is difficult to motivate such large cross-sections in standard $\Lambda$CDM cosmology, and one needs to invoke further exotic scenarios. 

A couple of additional comments are in order. One might imagine that the boosted \cnub{} component may show up in a detector like PTOLEMY\,\cite{PTOLEMY:2019hkd}. However, the event rate in PTOLEMY from such a boosted component is tiny, and hence not observable. Secondly, the upscattering of the \cnub{} can also take place from high energy cosmogenic neutrinos, coming from far-away sources such as active galactic nuclei~\cite{PhysRevLett.49.234}. The resultant boosted \cnub{} component can be an important source of flux at higher energies, and have consequences for neutrino telescopes. One can also imagine similar boosting mechanisms to take place from solar and atmospheric neutrinos, however these sources are rather nearby, and hence boosting may not be efficient. 

\section*{Acknowledgements}
AD was supported by the U.S. Department of Energy under contract number DE-AC02-76SF00515.
This work has made use of the Hamilton HPC Service of Durham University.

\bibliography{sinu}
\bibliographystyle{apsrev4-1}
\end{document}